\documentclass{aastex}

\usepackage{graphicx}
\usepackage{amsmath}
\usepackage{amssymb}
\usepackage{amsfonts}
\usepackage{amsthm}
\usepackage{color}

\usepackage[utf8]{inputenc}
\usepackage{fontenc}
\usepackage{bm}

\usepackage{natbib}

\newcommand{\Rsun}{\,$\textrm{R}_{\odot}$}
\newcommand{\Ha}{H$\alpha$}
\newcommand{\Ang}{\,\textrm{\AA}}
\newcommand{\kmps}{\,\mbox{km\,s$^{-1}$}}
\newcommand{\degree}{$^{\circ}$}

\newcommand{\ignore}[1]{}

\begin{document}

\title{Pre-eruption Oscillations in Thin and Long Features in a Quiescent Filament}

\author{Anand D. Joshi\altaffilmark{1,2}, Yoichiro Hanaoka\altaffilmark{1}, Yoshinori Suematsu\altaffilmark{1}, Satoshi Morita\altaffilmark{1}}
\affil{$^{1}$National Astronomical Observatory of Japan, Mitaka, Tokyo 181-8588, Japan.}
\affil{$^{2}$Korea Astronomy and Space Science Institute, Daejeon 34055, Republic of Korea.}

\author{Vasyl Yurchyshyn\altaffilmark{2,3}}
\affil{$^{3}$Big Bear Solar Observatory, New Jersey Institute of Technology, California, USA.}

\and

\author{Kyung-Suk Cho\altaffilmark{2,4}}
\affil{$^{4}$University of Science and Technology, Daejeon 34113, Republic of Korea.}

\begin{abstract}
We investigate the eruption of a quiescent filament located close to an active region. Large-scale activation was observed in only half of the filament in the form of pre-eruption oscillations. Consequently only this half erupted nearly 30\,hr after the oscillations commenced. Time-slice diagrams of 171\Ang\ images from Atmospheric Imaging Assembly were used to study the oscillations. The oscillations were observed in several thin and long features connecting the filament spine to the chromosphere below. This study traces the origin of such features and proposes their possible interpretation. Small-scale magnetic flux cancellation accompanied by a brightening was observed at the footpoint of the features shortly before their appearance, in images recorded by the Helioseismic and Magnetic Imager. Slow rise of the filament was detected in addition to the oscillations, indicating a gradual loss of equilibrium. Our analysis indicates that a change in magnetic field connectivity between two neighbouring active regions and the quiescent filament resulted in weakening of the overlying arcade of the filament leading to its eruption. It is also suggested that the oscillating features are filament barbs, and the oscillations are a manifestation during the pre-eruption phase of filaments.
\end{abstract}


\section{Introduction}\label{Sintro}

Solar filaments are plasma suspended in the corona above the photospheric polarity inversion line (PIL) \citep{Martin1998a}, and are composed of material that is cooler and denser than the surrounding corona \citep{Tandberg-Hanssen1995}. Although a filament may not appear to change much during a temporal scale of several hours, over the years, observations with improved resolution have gradually revealed the dynamic nature of filaments at the smallest scales \citep{Zirker.etal1998,Engvold.etal2001,Lin.etal2005a,Ning.etal2009,Vial.etal2012}. Eruptions of filaments are accompanied by coronal mass ejections \citep[CMEs,][]{Munro.etal1979} and flares \citep{Hirayama1974}. It is now well established that the different components of an eruption like flare ribbons, erupting filaments, CMEs, Moreton waves, etc. are manifestations of restructuring of the coronal magnetic field accompanied by the release of free energy \citep{Low1996,Priest.Forbes2002}.

However, before an eruption occurs, several events or features are observed, when the filament is said to be in an activated state. \citet{Sterling.etal2001} and \citet{Joshi.Srivastava2011a} have reported a slow rise phase of a filament, with rising speed of $1-10$\,\kmps\ before a rapid eruption, which typically occurs at speeds over 100\,\kmps. The slow-rise phase indicates onset of tether-cutting reconnection \citep{Forbes.Isenberg1991} beneath the filament, and the rapid eruptions occurs when the overlying coronal arcade catastrophically gives way to the rising filament \citep{Sterling.Moore2005}. Another pre-eruption phenomenon reported by \citet{Isobe.Tripathi2006} and \citet{Isobe.etal2007} is the oscillations of filaments, which were observed during the slow rise phase of the filament. Since the velocity amplitude of these oscillations exceeded 20\,\kmps, they are referred to as large-amplitude oscillations. Such oscillations are also known to be the result of nearby flares or transient events \citep{Jing.etal2003,Okamoto.etal2004}. On the other hand, small-amplitude oscillations in filaments, which have velocity amplitude $\sim$\,3\,\kmps, are also reported by \citet{Oliver.Ballester2002} and \citet{Soler.etal2009b}, which may or may not be linked to the filament eruption. \citet{Lin.etal2009} observed similar small-amplitude transverse oscillations in individual filament threads, which were interpreted as the kink modes of magnetohydrodynamic (MHD) waves.

\citet{Su.etal2012a} reported rotating long and narrow features, called ``tornadoes'' underneath three limb prominences observed before eruption. In one of these events, the footpoint of such a feature showed vortical magnetic field in the photosphere, leading the authors to suggest that photospheric vortex flows induced a twist in magnetic flux tubes giving rise to the tornadoes. \citet{Li.etal2012} however suggested that the rotating motions in a tornado could be a result of helical flows along the magnetic field in the cavity above the prominence, when viewed from one end of the field. The same event was also analysed by \citet{Panesar.etal2013} who observed that the tornadoes were triggered by three M-class flares occurring the nearby active region. However, \citet{Wedemeyer.etal2013} reported that the tornadoes might simply be legs of prominences. \citet{Panasenco.etal2014} also hold the view that counter-streaming flows and/or oscillatory motions in the filament may lead to the ``vortical illusion'' of a tornado.

In this study, we analyse activation and subsequent eruption of a large quiescent filament. The filament was situated close to the west limb in southern hemisphere. NOAA active region (AR) 11817 was located to the north-east of the filament. Two flares occurred in the AR shortly before the filament eruption\ignore{ generated a coronal wave that might have triggered the eruption}. Upon observing the filament for a sufficiently long time, we find that it was exhibiting activation in the form of pre-eruption oscillations. These oscillating features appeared thin and long, like barbs, connecting the filament to the low chromosphere. We investigate the filament oscillations, formation of the oscillating features, and their role in the eventual eruption of the filament.

\section{Observations}

\begin{figure}[!t] \begin{center}
\plotone{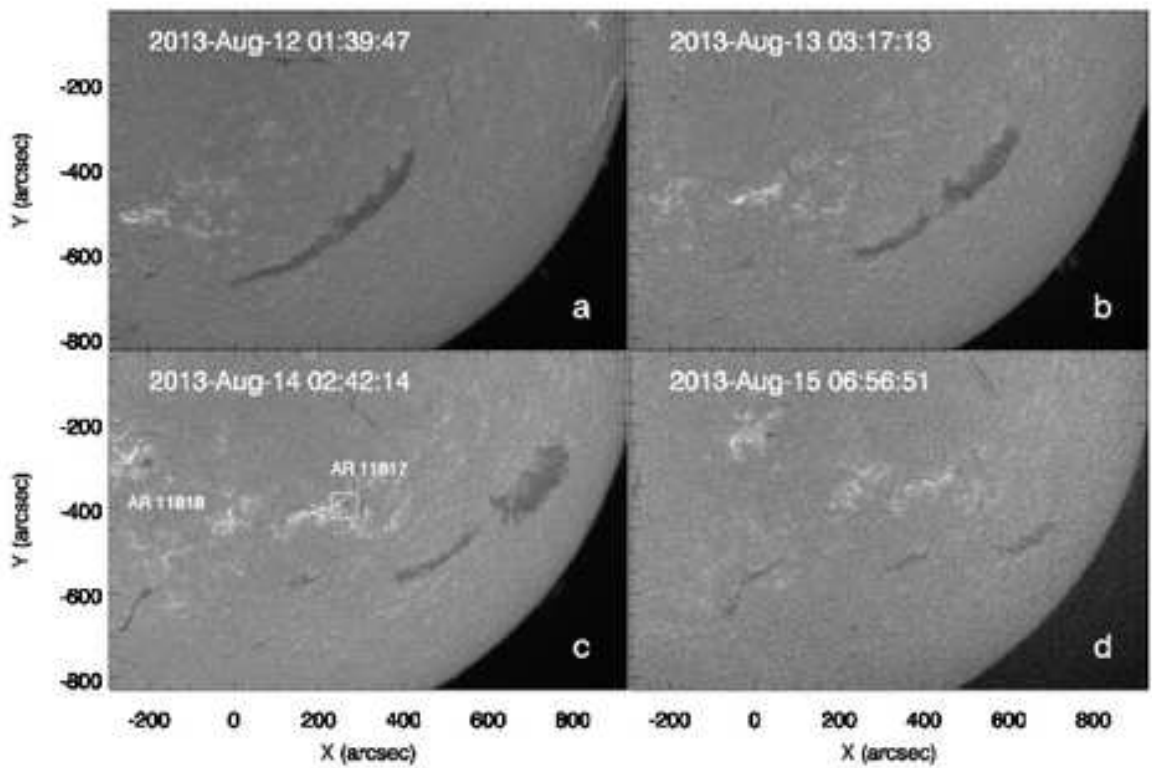}
\caption{\Ha\ line centre images from Solar Flare Telescope at National Astronomical Observatory of Japan. The box in panel (c) denotes the field-of-view observed by the Visible Imaging Spectrometer, shown in Figure~\ref{F:VISmosa}. Active regions 11817 and 11818 are also marked panel (c).}
\label{F:SFTmosa}
\end{center} \end{figure}

A large quiescent filament located near the south-west limb of the Sun erupted on 14 August 2013. \Ha\ images taken by the Solar Flare Telescope (SFT) at National Astronomical Observatory of Japan (NAOJ) \citep{Sakurai.etal1995} can be seen in Figure~\ref{F:SFTmosa} from 12 to 15 August 2013. In addition to other wavelengths, SFT captures \Ha\ line centre images at a cadence of about 30\,s, as well as images in red and blue wings at $\pm$\,0.5\Ang, $\pm$\,0.8\Ang, and in the red wing continuum at +3.5\Ang\ at a reduced cadence. We find remarkable day-to-day change in the quiescent filament in these four days. On 12 August we observed a single large filament. It appeared to be undergoing a separation roughly at the middle on 13 August. On 14 August, a few hours before the eruption, we can see the two halves completely separated in \Ha\ line centre. The eastern segment more or less retained its original shape, but the western segment appeared broader, indicating that it has either accumulated additional mass or risen in height, or both. Finally after the eruption, we can only see the eastern segment on 15 August. The change in apparent length however must be attributed to its proximity to the west limb of the Sun, i.e. foreshortening, and not actual change of its physical length. A large coronal mass ejection (CME) was associated with the eruption of the quiescent filament, which was first observed at 21:24\,UT on 14 August 2013 in the Large Angle Spectrometric Coronagraph on the Solar and Heliospheric Observatory (SoHO/LASCO) \citep{Domingo.etal1995b,Brueckner.etal1995}.

\begin{figure}[!t] \begin{center}
\includegraphics[width=0.52\textwidth,clip=]{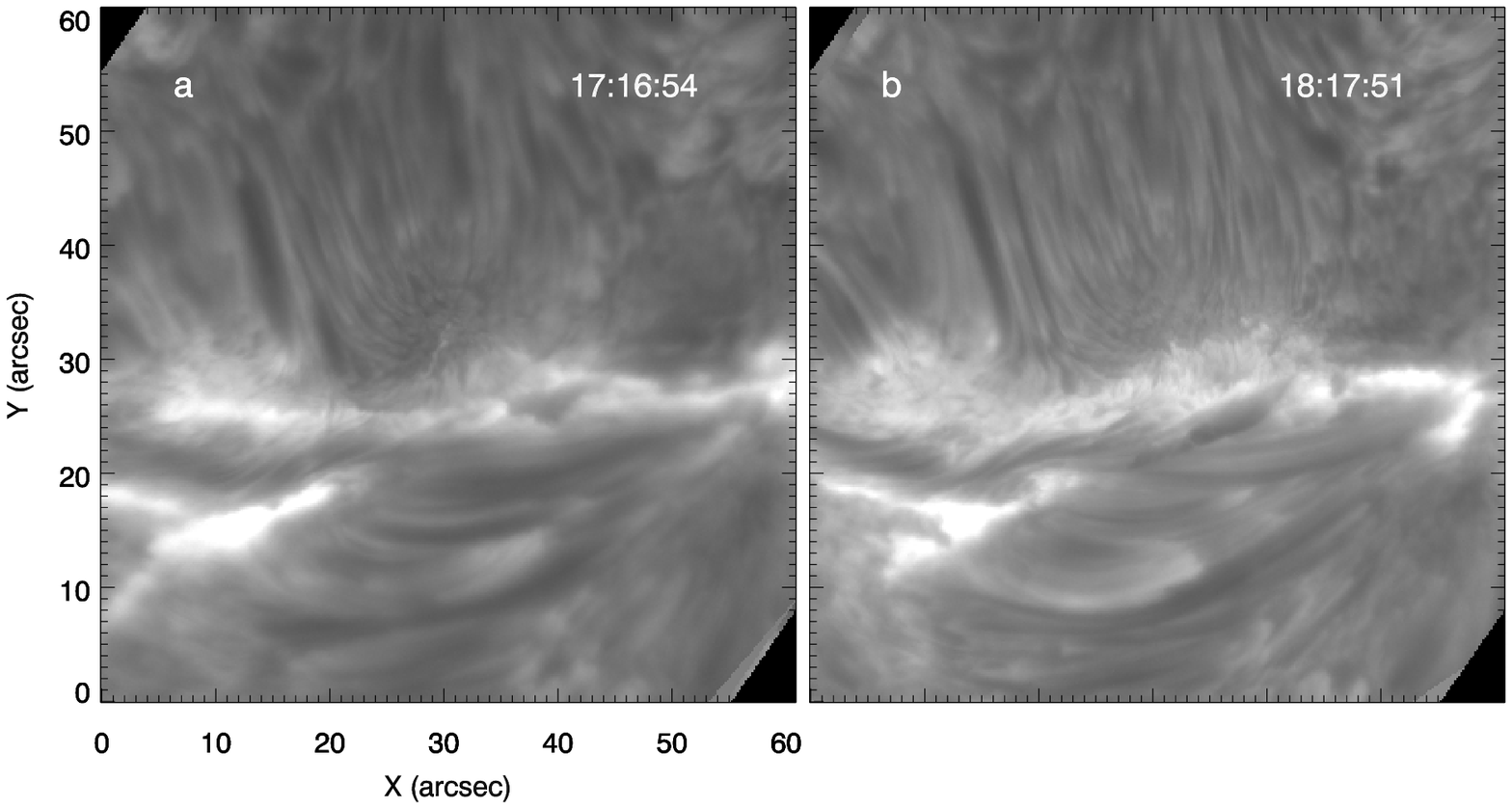}
\includegraphics[width=0.40\textwidth,clip=]{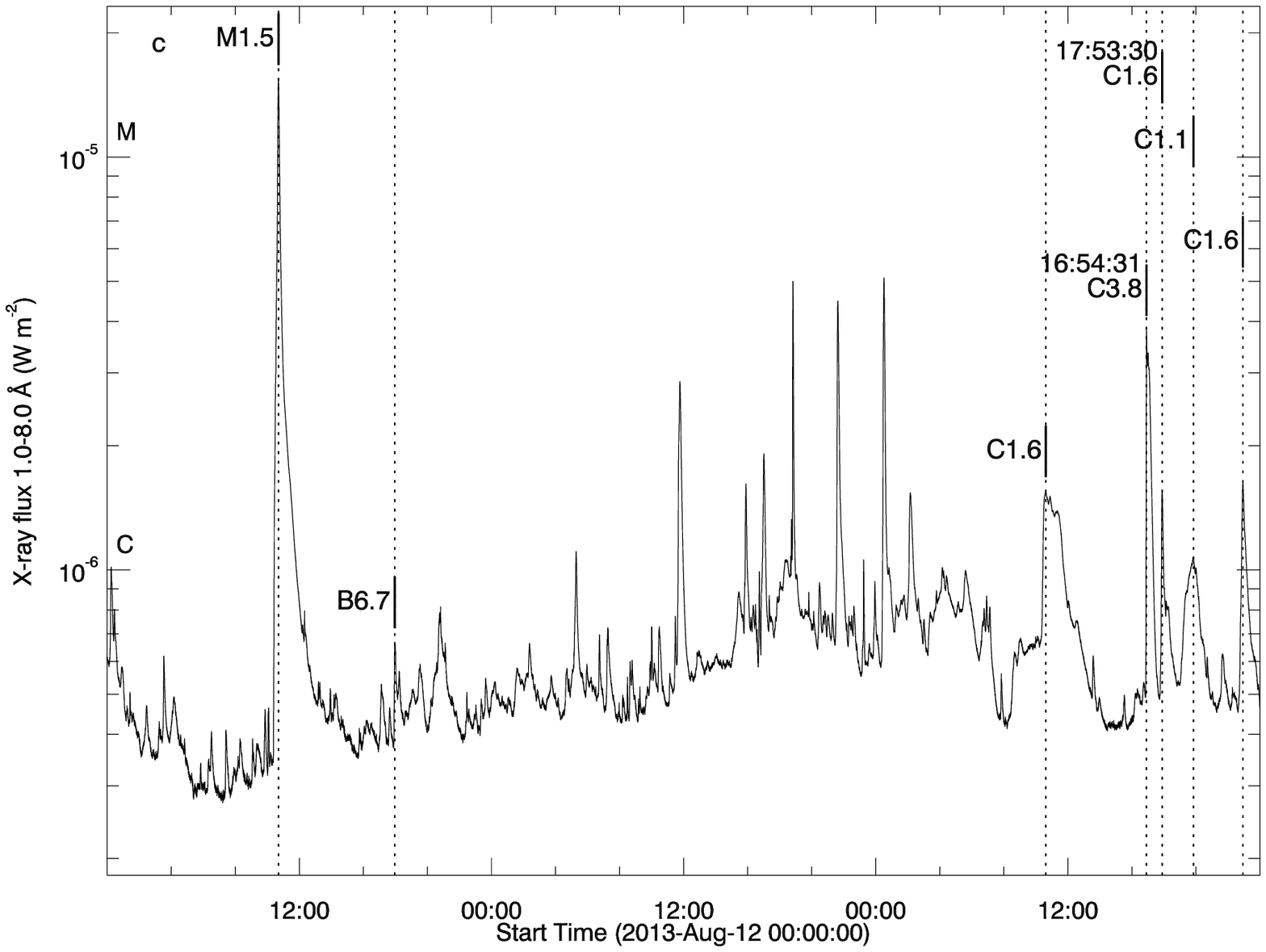}
\caption{\Ha\ images during the decay phase of the flares on 2013 August 14 observed by VIS are shown in panels (a) and (b). Panel (c) shows the soft X-ray flux from GOES from 00:00\,UT on 12 August 2013 to 24:00\,UT on 14 August 2013. Vertical dotted lines denote the flares occurring in AR\,11817 along with the X-ray class. The two flares peaking at 16:54:31\,UT and 17:53:30\,UT occurred just before the filament eruption.}
\label{F:VISmosa}
\end{center}\end{figure}

AR\,11817 was located to the north-east of the quiescent filament, at 34S 18W on 14 August 2013. The chromospheric plages are visible in Figure~\ref{F:SFTmosa}. This AR was observed by the Visible Imaging Spectrometer (VIS) on the New Solar Telescope at Big Bear Solar Observatory (BBSO/NST) from 17:08\,UT to 20:14\,UT, with an approximate cadence of 17\,s. An AR filament was located in the VIS field-of-view (FOV) to the south of the sunspot. Two C-class flares were recorded from AR\,11817 during this period, shortly before the eruption of the quiescent filament. The GOES soft X-ray class and peak times of the flares were C3.8 at 16:54\,UT and C1.6 at 17:53\,UT. Only the decay phase of the flare at 16:54\,UT was observed by VIS. Previously, an M1.5 flare at 10:42\,UT\ on 12 August and a C1.6 flare at 10:37\,UT on 14 August had occurred in this region. The AR showed frequent brightenings during the observation period on 14 August. Figure~\ref{F:VISmosa} shows two images of the AR observed by VIS on 14 August, and also the soft X-ray flux recorded by GOES satellite. The flares occurring in AR\,11817 are marked using vertical dotted lines at their peak times. The two flares which occurred just before the eruption of the quiescent filament are also shown along with their peak times in the figure.

The temporal and spatial proximity of the quiescent filament eruption to the flares in the AR suggests that the former was triggered by the flares. To verify this we searched for transient features, such as a coronal wave that could trigger the eruption, in running difference of 171\Ang\ and 304\Ang\ images from Atmospheric Imaging Assembly on board the Solar Dynamics Observatory (SDO/AIA) \citep{Lemen.etal2012} from 16:55\,UT to 20:25\,UT on 14 August 2013. The two wavelengths show the corona at 0.7\,MK and 80,000\,K, respectively. However, no such feature was observed even after varying the time difference between the two subtracted images. To rule out the possibility of a wave visible only in high temperature, we also took running difference of the 131\Ang\ images, showing the corona at 10\,MK. Even at this temperature, we could observe no trigger.

The time-lapse AIA 171\Ang\ movie reveals continuous mass motion along the filament spine. Although SFT uses full-disc observations, this motion could be observed, particularly on 13 August, before the filament is split into two distinct parts in \Ha\ line centre images. Furthermore, during the eruption, it is only the western segment that erupted outward, while the eastern segment remained more or less unchanged, as could be seen in the images on 2013 August 14 and 15 in Figure~\ref{F:SFTmosa}. We therefore investigate the filament activation and eruption phases in more detail.

\section{Analysis}

\subsection{Oscillations in the Filament}

\begin{center}
\begin{figure}[!htp]
\plotone{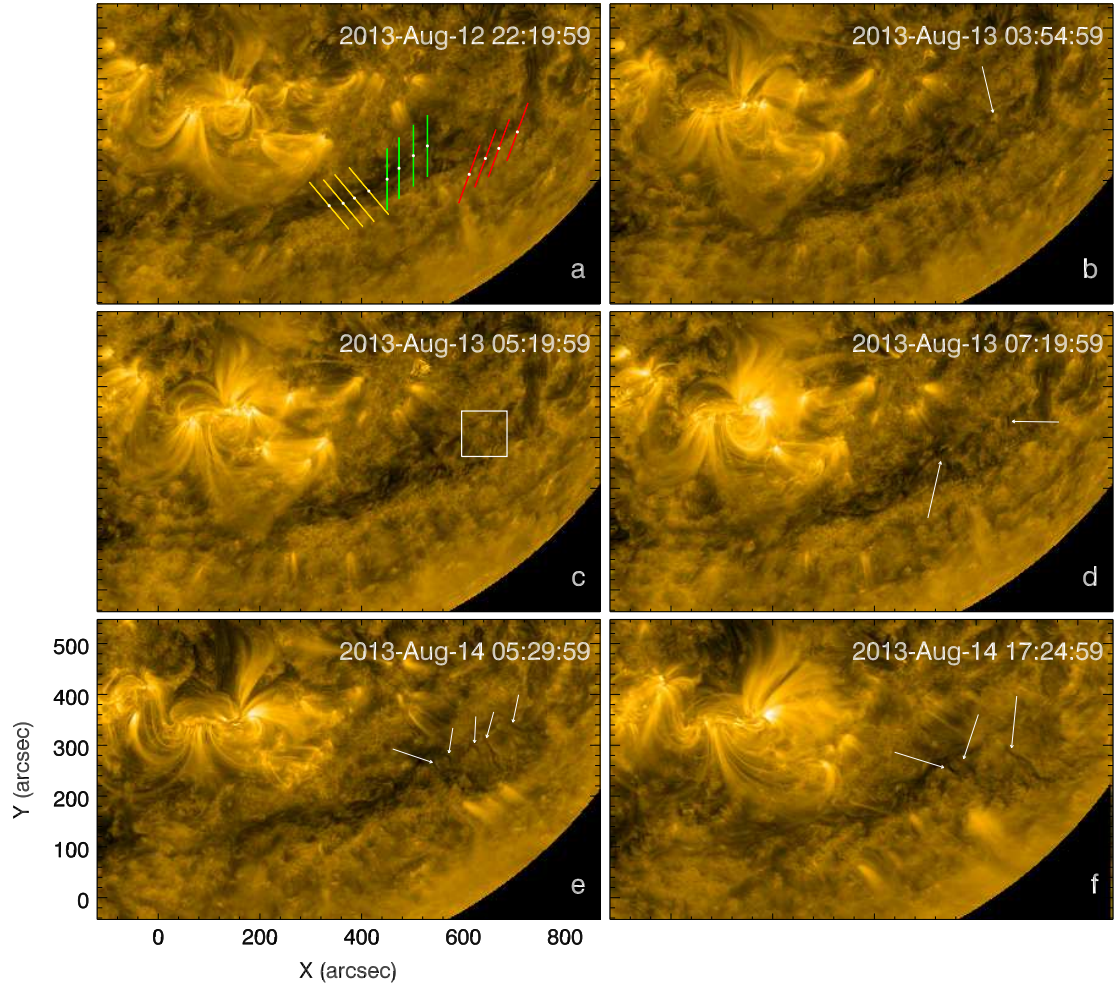}
\caption{Sequence of AIA 171\Ang\ images. Three sets of slits in (a) are used to extract time-slice diagrams. Arrows in panels (b) and (d) show brightening observed underneath the filament where subsequently oscillating features are observed, as indicated in panels (e) and (f).}
\label{F:171mosa}
\end{figure}
\end{center}

We used AIA 171\Ang\ images from 16:00\,UT on 12 August 2013 to 20:25\,UT on 14 August 2013, i.e. a period of more than 52\,hr, in order to study the pre-eruption activity in the quiescent filament. Interestingly enough, the filament spine appeared to be behaving differently on either side of a its midpoint. The eastern segment was relatively quiet, whereas in the western segment motion along the filament threads could be observed for several hours prior to the eruption. We make use of reduced cadence of 5\,min instead of the available 12\,s so that changes over a long period can be discerned in the time-lapse movie. Figure~\ref{F:171mosa} shows a sequence of six 171\Ang\ images from 12 to 14 August. For the sake of comparison all the images are derotated to a common time. An animation of the image sequence covering the entire time duration is provided in the online version of the paper. The quiescent filament stretches across the centre of the images.

The time-lapse movie reveals oscillatory behaviour in the spine. To confirm this we make use of time-slice diagrams. We selected a set of four points in the eastern segment, and two such sets in the western segment of the filament, as shown in Figure~\ref{F:171mosa}a. A slit, with a length of 200 pixels or 120\arcsec, was centred at each of these points, and intensity of the 171\Ang\ images is extracted along the slit. The set of slits in eastern segment are shown in yellow colour, while the two sets of slits in western segment are shown in green and red colours. To determine the direction along which the filament showed the maximum amplitude, orientation of the slits was varied from 0\degree\ to 175\degree\ with a 5\degree\ step, thereby generating 36 time-slice diagrams at each point \citep[cf.][]{Luna.etal2014}.

\begin{figure}[!t] \begin{center}
\includegraphics[width=0.95\textwidth,clip=]{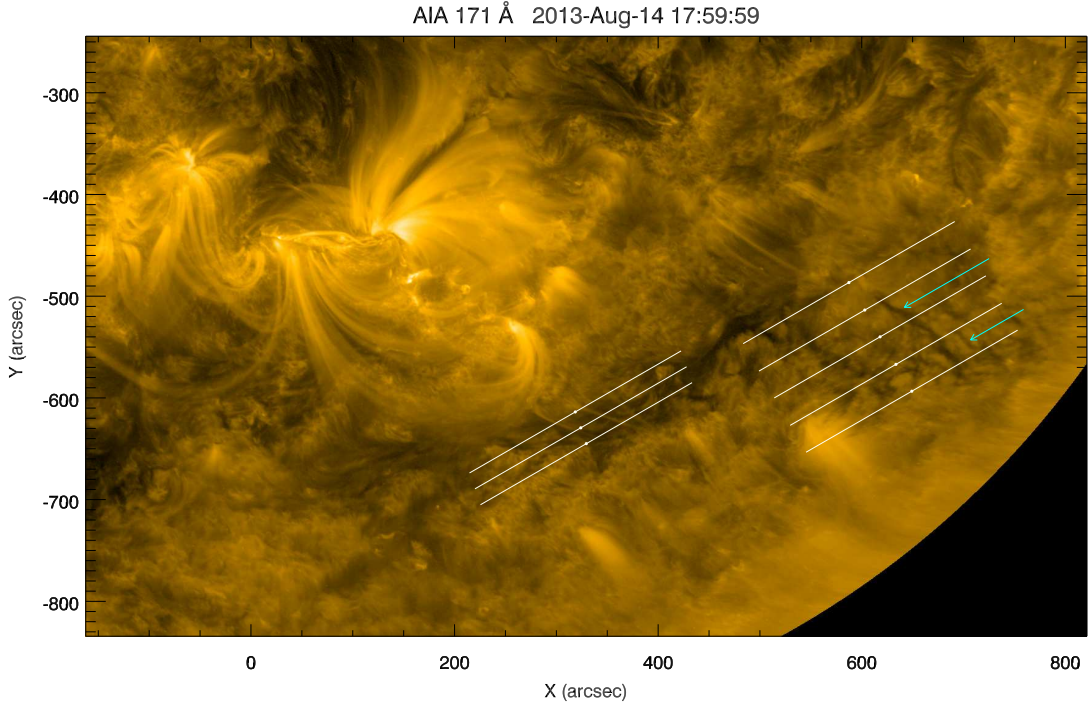}
\caption{Locations of two sets of parallel slits on AIA 171\Ang\ image oriented at an angle of 30\degree\ from the horizontal. Two completely formed oscillating features, as pointed out with the help of arrows, are visible in the western segment.}
\label{F:TSdia}
\end{center} \end{figure}

Upon inspecting the time-slice diagrams from each point at all the slit orientations, we find that the maximum amplitude occurs at an angle of 30\degree\ as measured from the horizontal direction, i.e. the maximum amplitude occurs along the filament spine. During the activation, the western segment of the filament rises steadily during 13 and 14 August. To capture this rise in the time-slice diagram, we selected five equally spaced slits in this part oriented at an angle of 30\degree\ to the horizontal. To compare motion in the two segments, three more such slits were selected in the eastern segment with the same orientation. Both these sets are 400 pixels (240\arcsec) long, and can be seen in Figure~\ref{F:TSdia}.

In Figure~\ref{F:171mosa}b and \ref{F:171mosa}d, we observe small-scale brightenings, as pointed out using arrows. Subsequently, thin narrow structures appear extending from the location of these brightenings toward the filament spine. Several such structures are marked using arrows in Figure~\ref{F:171mosa}e and \ref{F:171mosa}f. The western-most structure in panel (f) is the most distinct of all those observed. The white box in Figure~\ref{F:171mosa}c marks the area around this structure, which will be discussed later.

\begin{figure}[!t] \begin{center}
\plotone{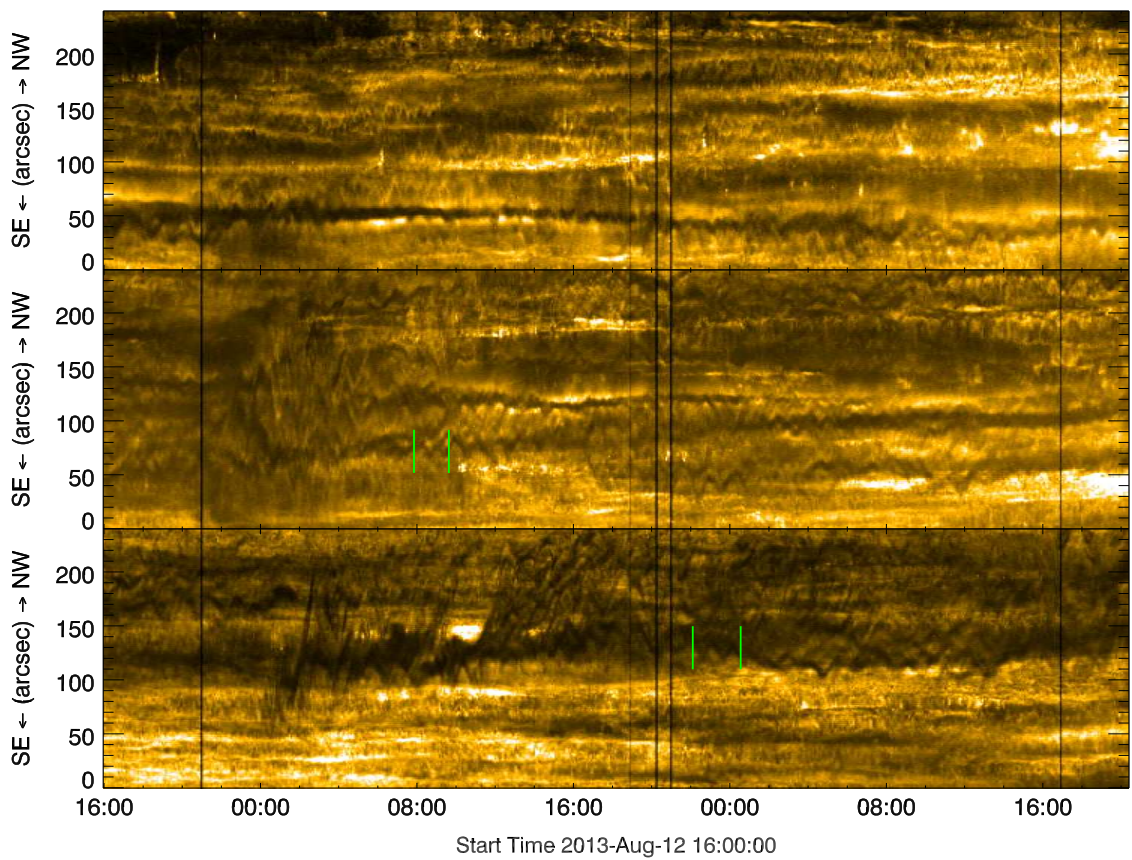}
\caption{Time-slice diagrams along the three slits in the eastern segment of the filament, as seen in Figure~\ref{F:TSdia}. Y-axis shows distance in arcsec where NW and SE indicate the directions north-west and south-east. Two pairs of vertical lines in middle and bottom panels mark the time interval of two of the observed oscillating patterns.}
\label{F:TSoscE}
\end{center} \end{figure}

\subsubsection{The Eastern Segment}

Figure~\ref{F:TSoscE} shows the time-distance plots along the three slits in the eastern segment of the quiescent filament. As one goes from top to bottom panel, the slits shift away from the disc centre towards the solar limb. Any motion of plasma along the slit would be visible as streaks or a pattern in the vertical direction. The top panel which shows the slit farthest from the limb shows no such motion along the slit. The middle panel, represents the central slit and shows a few oscillatory patterns. The notable ones are between 08:00 and 16:00\,UT\ on 13 August, and between 23:00\,UT\ on 13 August and 08:00\,UT\ on 14 August. Patterns in these time intervals can be seen between 50\arcsec\ and 150\arcsec. All these oscillatory patterns could be observed for about 2-4 cycles only, and their amplitudes appear constant.

The bottom panel shows more of such oscillatory patterns and for a longer duration than the middle panel. They can be discerned from 00:00\,UT on 13 August to 06:00\,UT on 14 August, i.e. for a period of more than a full day. In addition, we can see several dark streaks with a positive slope from about 11:00\,UT\ to 17:00\,UT\ on 13 August, indicating material flowing in the north-west direction, i.e. towards the western segment of the filament. It is possible to directly observe this mass flow in 171\Ang\ time-lapse movie as well, but it becomes more evident in the time-slice diagram.

\subsubsection{The Western Segment}

\begin{figure}[!t] \begin{center}
\epsscale{0.80}
\plotone{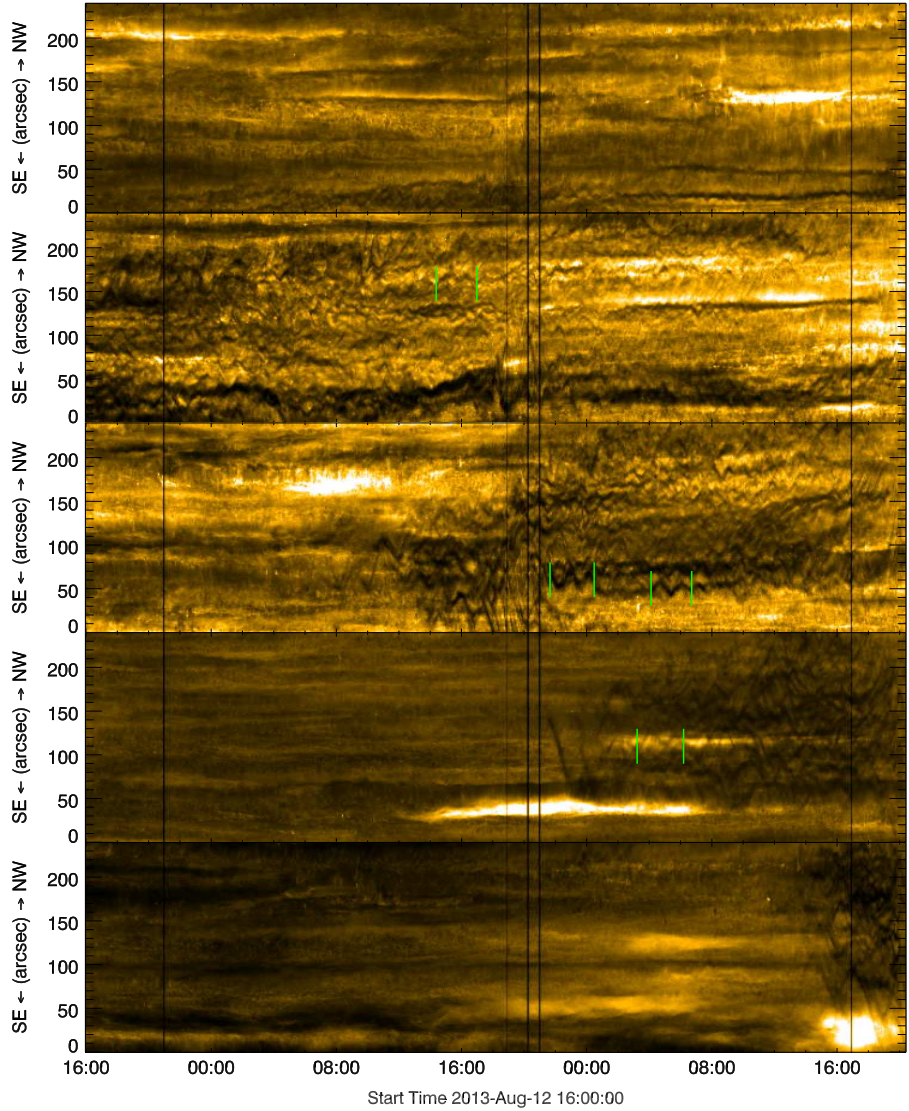}	
\caption{Time-slice diagrams along the five slits in the western segment of the filament, as seen in Figure~\ref{F:TSdia}. Y-axis shows distance in arcsec and the arrows pointing to NW and SE indicate the directions north-west and south-east. The four pairs of vertical lines in three panels mark the time interval in which four of the oscillating patterns were observed.}
\label{F:TSoscW}
\end{center} \end{figure}

Figure~\ref{F:TSoscW} shows time-distance plots along the five slits in the western segment of the quiescent filament. The panels from top to bottom show the slits from disc centre towards the solar limb. As before, the top panel representing the slit farthest from the limb shows almost no noticeable motion. The second panel shows several patterns from the beginning of the observation period, i.e. 16:00\,UT\ on 12 August until 00:00\,UT\ on 14 August, especially between 150\arcsec\ and 200\arcsec\ on the vertical axis.

We can see the oscillations most clearly in the middle panel of Figure~\ref{F:TSoscW}, from about 12:00\,UT\ on 13 August to 12:00\,UT\ on 14 August. Oscillations in the bottom two panels in Figure~\ref{F:TSoscW} are observed only on 14 August. In fact, right from the middle panel to the bottom panel, oscillations sequentially appear at later times. This indicates a slow rise of the quiescent filament, whereby the oscillating structure crosses the slit. From these three panels, we may conclude that the slow-rise of the filament commenced at around 12:00\,UT on 13 August, i.e. more than 30\,hr before the eruption.

The arrows in Figure~\ref{F:171mosa}e and \ref{F:171mosa}f point to the base of the oscillating features. These appear similar to the tornadoes that were previously reported by many researchers. The tornadoes exhibit apparent rotational or vortical motion, and are most clearly observed at the limb between a prominence spine and low chromosphere \citep{Su.etal2012a}. However, owing to a lack of Doppler shift observations with sufficient spatial resolution, we do not find evidence of rotational motion in the structures. When we focus on the location where these features are rooted in the low chromosphere in AIA 171\Ang\ images, we find instances of brightenings at these points, possibly indicating plasma heating. Endpoint brightenings were reported by \citet{Wang.etal2009} at the outer edges of filament channels, indicating flux cancellation. \citet{Wang.Muglach2013} observed bi-directional jets after such events, whereas, in our case we saw the development of a feature connected to the spine above at the location of brightening.

\cite{Chae.etal2000} have also detected magnetic flux cancellation followed by injection of mass into a filament. The broadening of the filament in Figure~\ref{F:SFTmosa}c could be thus due to an increase in mass, injected into the filament by the oscillating features. From the animation of Figure~\ref{F:171mosa}, we find that the filament began to rise rapidly on 14 August. Therefore, as suggested earlier, the broadening of the western segment can now be attributed to both, mass accumulation \citep{Chae.etal2000} as well as overall slow rise.

The maximum amplitude of the oscillations is observed when the slit is oriented along the filament spine. Thus, we observe transverse oscillations in thin and long structures oriented almost perpendicular to filament spine, over more than 30\,hr before the eruption. Through the extracted oscillations, we find that period ranges from 60 to 90\,min, with a velocity of about 3-5\,kmps\ in the plane of sky. Several patterns are observed at different locations along a given slit, implying the presence of more than one oscillating structures, however, not with the same phase. Another important aspect is the drift in the oscillations. We do not see any drifting patterns or streaks in the western segment unlike in Figure~\ref{F:TSoscE}, indicating that the filament mass in the western segment does not flow along the filament spine.

\subsection{Barb Footpoint}

\begin{figure}[!ht] \begin{center}
\plotone{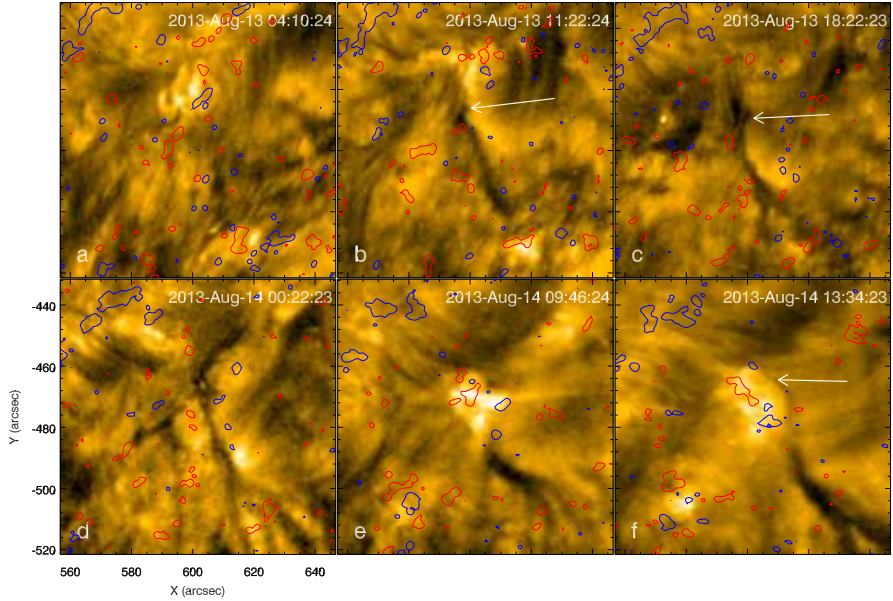}
\caption{AIA 171\Ang\ images of the area outlined by the white box shown in Figure~\ref{F:171mosa}c. Blue and red contours show respectively positive and negative polarities of HMI magnetograms at $\pm$\,20\,G level. Thin and narrow structures are seen in panels (b), (c) , (d) and (f) rooted at the location of brightenings in panels (a) and (e). Magnetic fields of both polarities are visible at the location of brightenings in panels (a), (e) and (f).}
\label{F:tornado}
\end{center} \end{figure}

To trace the origin of the oscillating barbs, we focus on the box shown in Figure~\ref{F:171mosa}c. Figure~\ref{F:tornado} shows AIA 171\Ang\ images of this box at six different times. HMI magnetic field contours at $+20$\,G (blue) and $-20$\,G (red) levels are overplotted on these images. In panel (a) of the figure, we observe a brightening near the centre of the image, accompanied by magnetic field concentrations of both polarities. As the brightening disappears, both polarities too are seen to disappear. Such small-scale brightenings in EUV images are reported to be associated with flux cancellation along a filament channel by \citet{Wang.Muglach2013}. They have found such brightenings to last for a period of no more than 10\,min. \citet{Chae.etal2001} have reported emergence of magnetic bipoles in a filament channel, while \citet{Chae2003} have found jets accompanying small-scale cancellation of magnetic fields, which lasted for up to 30\,min. On the other hand, \citet{Joshi.etal2013} found mass flow in filament spine responsible for the formation of a barb.

Soon after the first appearance of the brightening, we find a dark long and narrow structure rooted at the same location \citep{Chae.etal2000}, shown in panel (b). As the filament rises, this structure is seen to elongate, and resembles a filament barb. Later, on 14 August, in panel (e), we can once again see the brightening accompanied by the appearance of both polarities. Notice that in the field shown in this figure, positions of both the brightening as well as the root of the structure are seen to shift together. This further strengthens our claim that the brightening indeed occurred at the footpoint of this structure. In panel (f), the brightness appears over a smaller area than in panel (e), however, the footpoint appears very thin, indicating that the structure is extended well beyond the field-of-view shown here, as can confirmed from Figure~\ref{F:171mosa}f.

\subsection{Magnetic Field Extrapolation}

\begin{figure}[!t] \begin{center}
\plotone{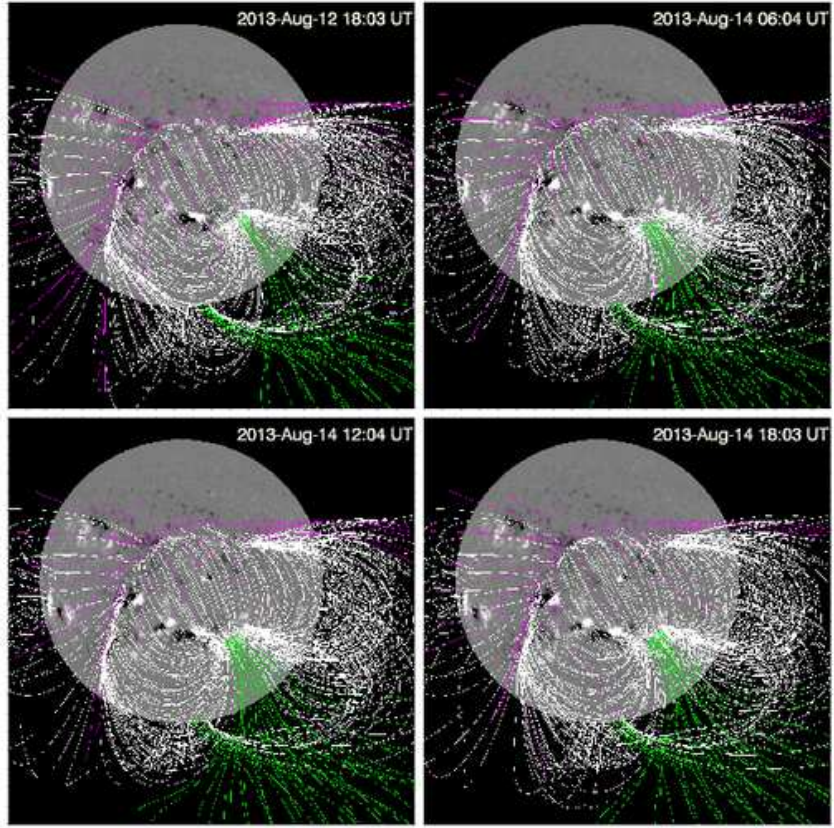}
\caption{Potential field source surface model showing global magnetic field connectivity. AR~11817 is located near the disc centre and AR~11818 is visible to its north-east. The green lines to the west of AR~11817 indicate open magnetic field lines that are rooted close to the quiescent filament.}
\label{F:PFSS}
\end{center} \end{figure}

We carried out a potential field source surface (PFSS) extrapolation \citep{Schatten.etal1969,Schrijver.DeRosa2003} using the SolarSoft package \texttt{pfss}. Results from PFSS are shown in Figure~\ref{F:PFSS}. The solar disc shows photospheric magnetic field saturated at $\pm$\,100\,G. The potential field is plotted from the solar photosphere up to a radius of 2.5\Rsun. White lines represent closed field lines connecting opposite magnetic field regions, while red and green lines represent open magnetic field lines originating from negative and positive regions respectively, and crossing the upper boundary\ignore{of 2.5\Rsun} of the extrapolation domain.

AR\,11817 is located near the disc centre while AR\,11818 is visible farther east near the equator. We observe that field lines connect a small region to the west of AR\,11817 to AR\,11818 and to other regions on the solar disc. Open field lines are originating from positive fields in the region to the west of AR\,11817 are shown in green colour. The quiescent filament under study is located to the south-west of AR\,11817.

If we accept the typical height at which a filament is suspended in the solar corona to be about 10,000\,km, then the overlying field lines shown in Figure~\ref{F:PFSS} are at a much larger height. However, we find that compared to 18:03\,UT on 12 August, the footpoints of the field lines rooted near AR\,11817 shift gradually towards the AR itself, and away from the location of the quiescent filament. The difference in footpoint location appears most noticeable between the two images in the top row. We suggest that the footpoint shift has contributed to the release of the coronal arcade field enveloping the filament, causing it to undergo slow rise and eventually leading to the eruption.

\section{Results and Discussion}

We analyse the activation and eruption process of a quiescent filament on 14 August 2013 using 171\Ang\ images from AIA/SDO. The occurrence of two flares in the neighbouring AR 11817 just before the eruption was only coincidental, and we find that they did not directly affect the filament eruption. Only the western segment of the filament erupted, while the eastern segment remained more or less unchanged throughout the 52\,hr period of filament observations. The western segment showed activation in the form of oscillations in thin and long features, that resembled barbs, more than a day before the eruption commenced. We therefore constructed time-slice diagrams along the spine of the filament in both segments of the filament.

Figure~\ref{F:TSoscE} indicates mass flow from eastern to western segment for almost six hours, one day prior to the eruption. This is likely to have made the western segment unstable, reinforcing the oscillations. \ignore{We detect presence of oscillations in the}The time-lapse movie shows oscillating features in the filament spine on 13 August 2013, as seen in Figure~\ref{F:TSoscW}. During this time, the filament also undergoes a slow rise, which is a recognised form of pre-eruption activation. As the filament rises, we find oscillating structures connecting the spine to the chromosphere below. The important thing to note is that these structures can be traced back to a few tens of hours before the actual prominence eruption.
These structures resemble the ``tornadoes'' seen in Figure~1 from \citet{Su.etal2012a} and Figure~4 from \citet{Panasenco.etal2014}.

Dopplergrams were constructed from \Ha\,$\pm$\,0.5\,\Ang\ and \Ha\,$\pm$\,0.8\,\Ang\ using full-disc images acquired by SFT. However, the fine oscillating structures could not be resolved sufficiently. Since the oscillations were observed in thin structures, and not the entire filament, it was not possible to observe the ``winking'' of the filament in \Ha\ line centre or wing images either \citep{Okamoto.etal2004}. Without Doppler observations of sufficient resolution, or data from any of the two STEREO spacecraft, it is not possible to ascertain the vortical nature of these flows. Therefore, we use the broader and more conventional term to describe them, i.e. barbs.

For more than 24\,hr before the eruption, we observed activation in the form of a slow rise as well as  oscillations in the filament. The eastern segment showed oscillations only for a short duration, as seen in the bottom panel of Figure~\ref{F:TSoscE}. Whereas, the oscillations in the western segment commenced at sequentially later times as the filament rose in height, as shown from the third panel to the bottom panel in Figure~\ref{F:TSoscW}. Each oscillation event in a feature persisted for 2 to 4 periods, with each period between 60 and 90\,min, and the velocity amplitude in the range 3 to 5\,\kmps. 

The oscillations were transverse to the length of the barbs. Such oscillations are identified to be  kink MHD modes \citep{Edwin.Roberts1983} in thin threads of filaments. These modes are regarded as propagating waves with an average period of 3.6\,min in \citet{Lin.etal2009}. On the other hand, in a numerical study \citet{Soler.etal2010} regarded these modes as standing waves, with an approximate period of 23\,min. The observed periods in our case are much longer than in either scenario. A damping time of 2 to 4 periods for the kink modes strongly implies resonant absorption of the waves \citep{Soler.etal2009a}. However, we did not detect a significant decrease in amplitude in the duration over which a given oscillating feature was observed. Instead, the oscillations appeared to end all of a sudden. This could imply a change in the oscillating feature owing to the evolving solar filament, which renders a given thread indiscernible after a few cycles or
a mechanism with a very short damping time.

In the previous studies, whether observational or numerical, the threads are between 0.1\arcsec\ and 0.6\arcsec\ wide, while the barbs in this study are 4\arcsec--5\arcsec\ wide in many instances (Figure~\ref{F:tornado}), and the thin tube approximation may not be applicable here. In addition, previous researchers have observed the threads oriented along the filament spine, or modelled with both their ends rooted in the photosphere. However in the present study, one end of the barbs appears to be connected to the filament spine, which was rising slowly over several hours. Therefore, more theoretical as well as observational studies are required before the nature of oscillations in barbs is understood.

Prior to the appearance of the oscillating barbs, we see opposite magnetic fields at their footpoints, accompanied by a brightening in the EUV wavelength \citep{Wang.etal2009}. The opposite polarities disappear soon, implying either photospheric reconnection or simply cancellation of flux. Shortly afterward, we see thin and dark oscillating structures, i.e. barbs, extending from the location of the brightening. This was observed twice during the entire period of observations from two locations very close to one another, as seen in Figure~\ref{F:tornado}. Similar brightening preceded the appearance of at least one other barb as pointed in panels (b) and (d) of Figure~\ref{F:171mosa}. Such reconnecting regions near a filament spine have previously been found to be responsible for upflows in a filament barb \citep{Litvinenko2000}, as well as injection of mass \citep{Chae.etal2000}. In this case, we have observed the formation of the barb after the reconnection.

A potential field source surface extrapolation of the magnetic field connectivity between two active regions on the disc was carried out. Throughout the period of observations, a gradual change in the large scale magnetic field connecting the region near the quiescent filament and AR\,11818 was observed, which indicated a weakening of the field near the eastern side of the quiescent filament. This coupled with the activation in the form of oscillations and slow rise signified a steady loss of equilibrium of the quiescent filament, that eventually led to its eruption.

\section{Conclusions}

We report two simultaneous forms of pre-eruption activation of a quiescent filament, namely small-amplitude transverse oscillations in thin features and slow rise detected almost 30\,hr prior to the eruption. The oscillations occurred in thin and long features, and not along the entire filament spine. Small-scale flux cancellation was observed at the footpoint of these features prior to their appearance. These features appeared more prominent in the western segment of the filament than the eastern one. Subsequently, only the western segments underwent eruption, and a CME was seen to be associated with it. The oscillating features appeared thin and narrow, and they resembled tornadoes from some of the previous studies. The time-lapse movie shows that the filament, which appears tenuous in AIA 171\Ang\ as compared to \Ha\ images from the SFT, is anchored with the help of the oscillating structures, much like the way a barb connects the filament spine to the chromosphere below.


\begin{thebibliography}{}
\expandafter\ifx\csname natexlab\endcsname\relax\def\natexlab#1{#1}\fi

\bibitem[{{Brueckner} {et~al.}(1995){Brueckner}, {Howard}, {Koomen},
  {Korendyke}, {Michels}, {Moses}, {Socker}, {Dere}, {Lamy}, {Llebaria},
  {Bout}, {Schwenn}, {Simnett}, {Bedford}, \& {Eyles}}]{Brueckner.etal1995}
{Brueckner}, G.~E., {Howard}, R.~A., {Koomen}, M.~J., {et~al.} 1995, \solphys,
  162, 357

\bibitem[{{Chae}(2003)}]{Chae2003}
{Chae}, J. 2003, \apj, 584, 1084

\bibitem[{{Chae} {et~al.}(2000){Chae}, {Denker}, {Spirock}, {Wang}, \&
  {Goode}}]{Chae.etal2000}
{Chae}, J., {Denker}, C., {Spirock}, T.~J., {Wang}, H., \& {Goode}, P.~R. 2000,
  \solphys, 195, 333

\bibitem[{{Chae} {et~al.}(2001){Chae}, {Martin}, {Yun}, {Kim}, {Lee}, {Goode},
  {Spirock}, \& {Wang}}]{Chae.etal2001}
{Chae}, J., {Martin}, S.~F., {Yun}, H.~S., {et~al.} 2001, \apj, 548, 497

\bibitem[{{Domingo} {et~al.}(1995){Domingo}, {Fleck}, \&
  {Poland}}]{Domingo.etal1995b}
{Domingo}, V., {Fleck}, B., \& {Poland}, A.~I. 1995, \solphys, 162, 1

\bibitem[{{Edwin} \& {Roberts}(1983)}]{Edwin.Roberts1983}
{Edwin}, P.~M., \& {Roberts}, B. 1983, \solphys, 88, 179

\bibitem[{{Engvold} {et~al.}(2001){Engvold}, {Jakobsson}, {Tandberg-Hanssen},
  {Gurman}, \& {Moses}}]{Engvold.etal2001}
{Engvold}, O., {Jakobsson}, H., {Tandberg-Hanssen}, E., {Gurman}, J.~B., \&
  {Moses}, D. 2001, \solphys, 202, 293

\bibitem[{{Forbes} \& {Isenberg}(1991)}]{Forbes.Isenberg1991}
{Forbes}, T.~G., \& {Isenberg}, P.~A. 1991, \apj, 373, 294

\bibitem[{{Hirayama}(1974)}]{Hirayama1974}
{Hirayama}, T. 1974, \solphys, 34, 323

\bibitem[{{Isobe} \& {Tripathi}(2006)}]{Isobe.Tripathi2006}
{Isobe}, H., \& {Tripathi}, D. 2006, \aap, 449, L17

\bibitem[{{Isobe} {et~al.}(2007){Isobe}, {Tripathi}, {Asai}, \&
  {Jain}}]{Isobe.etal2007}
{Isobe}, H., {Tripathi}, D., {Asai}, A., \& {Jain}, R. 2007, \solphys, 246, 89

\bibitem[{{Jing} {et~al.}(2003){Jing}, {Lee}, {Spirock}, {Xu}, {Wang}, \&
  {Choe}}]{Jing.etal2003}
{Jing}, J., {Lee}, J., {Spirock}, T.~J., {et~al.} 2003, \apjl, 584, L103

\bibitem[{{Joshi} \& {Srivastava}(2011)}]{Joshi.Srivastava2011a}
{Joshi}, A.~D., \& {Srivastava}, N. 2011, \apj, 730, 104

\bibitem[{{Joshi} {et~al.}(2013){Joshi}, {Srivastava}, {Mathew}, \&
  {Martin}}]{Joshi.etal2013}
{Joshi}, A.~D., {Srivastava}, N., {Mathew}, S.~K., \& {Martin}, S.~F. 2013,
  \solphys, 288, 191

\bibitem[{{Lemen} {et~al.}(2012){Lemen}, {Title}, {Akin}, {Boerner}, {Chou},
  {Drake}, {Duncan}, {Edwards}, {Friedlaender}, {Heyman}, {Hurlburt}, {Katz},
  {Kushner}, {Levay}, {Lindgren}, {Mathur}, {McFeaters}, {Mitchell}, {Rehse},
  {Schrijver}, {Springer}, {Stern}, {Tarbell}, {Wuelser}, {Wolfson}, {Yanari},
  {Bookbinder}, {Cheimets}, {Caldwell}, {Deluca}, {Gates}, {Golub}, {Park},
  {Podgorski}, {Bush}, {Scherrer}, {Gummin}, {Smith}, {Auker}, {Jerram},
  {Pool}, {Soufli}, {Windt}, {Beardsley}, {Clapp}, {Lang}, \&
  {Waltham}}]{Lemen.etal2012}
{Lemen}, J.~R., {Title}, A.~M., {Akin}, D.~J., {et~al.} 2012, \solphys, 275, 17

\bibitem[{{Li} {et~al.}(2012){Li}, {Morgan}, {Leonard}, \&
  {Jeska}}]{Li.etal2012}
{Li}, X., {Morgan}, H., {Leonard}, D., \& {Jeska}, L. 2012, \apjl, 752, L22

\bibitem[{{Lin} {et~al.}(2005){Lin}, {Engvold}, {Rouppe van der Voort}, {Wiik},
  \& {Berger}}]{Lin.etal2005a}
{Lin}, Y., {Engvold}, O., {Rouppe van der Voort}, L., {Wiik}, J.~E., \&
  {Berger}, T.~E. 2005, \solphys, 226, 239

\bibitem[{{Lin} {et~al.}(2009){Lin}, {Soler}, {Engvold}, {Ballester},
  {Langangen}, {Oliver}, \& {Rouppe van der Voort}}]{Lin.etal2009}
{Lin}, Y., {Soler}, R., {Engvold}, O., {et~al.} 2009, \apj, 704, 870

\bibitem[{{Litvinenko}(2000)}]{Litvinenko2000}
{Litvinenko}, Y.~E. 2000, \solphys, 196, 369

\bibitem[{{Low}(1996)}]{Low1996}
{Low}, B.~C. 1996, \solphys, 167, 217

\bibitem[{{Luna} {et~al.}(2014){Luna}, {Knizhnik}, {Muglach}, {Karpen},
  {Gilbert}, {Kucera}, \& {Uritsky}}]{Luna.etal2014}
{Luna}, M., {Knizhnik}, K., {Muglach}, K., {et~al.} 2014, \apj, 785, 79

\bibitem[{{Martin}(1998)}]{Martin1998a}
{Martin}, S.~F. 1998, \solphys, 182, 107

\bibitem[{{Munro} {et~al.}(1979){Munro}, {Gosling}, {Hildner}, {MacQueen},
  {Poland}, \& {Ross}}]{Munro.etal1979}
{Munro}, R.~H., {Gosling}, J.~T., {Hildner}, E., {et~al.} 1979, \solphys, 61,
  201

\bibitem[{{Ning} {et~al.}(2009){Ning}, {Cao}, {Okamoto}, {Ichimoto}, \&
  {Qu}}]{Ning.etal2009}
{Ning}, Z., {Cao}, W., {Okamoto}, T.~J., {Ichimoto}, K., \& {Qu}, Z.~Q. 2009,
  \aap, 499, 595

\bibitem[{{Okamoto} {et~al.}(2004){Okamoto}, {Nakai}, {Keiyama}, {Narukage},
  {UeNo}, {Kitai}, {Kurokawa}, \& {Shibata}}]{Okamoto.etal2004}
{Okamoto}, T.~J., {Nakai}, H., {Keiyama}, A., {et~al.} 2004, \apj, 608, 1124

\bibitem[{{Oliver} \& {Ballester}(2002)}]{Oliver.Ballester2002}
{Oliver}, R., \& {Ballester}, J.~L. 2002, \solphys, 206, 45

\bibitem[{{Panasenco} {et~al.}(2014){Panasenco}, {Martin}, \&
  {Velli}}]{Panasenco.etal2014}
{Panasenco}, O., {Martin}, S.~F., \& {Velli}, M. 2014, \solphys, 289, 603

\bibitem[{{Panesar} {et~al.}(2013){Panesar}, {Innes}, {Tiwari}, \&
  {Low}}]{Panesar.etal2013}
{Panesar}, N.~K., {Innes}, D.~E., {Tiwari}, S.~K., \& {Low}, B.~C. 2013, \aap,
  549, A105

\bibitem[{{Priest} \& {Forbes}(2002)}]{Priest.Forbes2002}
{Priest}, E.~R., \& {Forbes}, T.~G. 2002, \aapr, 10, 313

\bibitem[{{Sakurai} {et~al.}(1995){Sakurai}, {Ichimoto}, {Nishino}, {Shinoda},
  {Noguchi}, {Hiei}, {Li}, {He}, {Mao}, {Lu}, {Ai}, {Zhao}, {Kawakami}, \&
  {Chae}}]{Sakurai.etal1995}
{Sakurai}, T., {Ichimoto}, K., {Nishino}, Y., {et~al.} 1995, \pasj, 47, 81

\bibitem[{{Schatten} {et~al.}(1969){Schatten}, {Wilcox}, \&
  {Ness}}]{Schatten.etal1969}
{Schatten}, K.~H., {Wilcox}, J.~M., \& {Ness}, N.~F. 1969, \solphys, 6, 442

\bibitem[{{Schrijver} \& {De Rosa}(2003)}]{Schrijver.DeRosa2003}
{Schrijver}, C.~J., \& {De Rosa}, M.~L. 2003, \solphys, 212, 165

\bibitem[{{Soler} {et~al.}(2010){Soler}, {Arregui}, {Oliver}, \&
  {Ballester}}]{Soler.etal2010}
{Soler}, R., {Arregui}, I., {Oliver}, R., \& {Ballester}, J.~L. 2010, \apj,
  722, 1778

\bibitem[{{Soler} {et~al.}(2009{\natexlab{a}}){Soler}, {Oliver}, \&
  {Ballester}}]{Soler.etal2009b}
{Soler}, R., {Oliver}, R., \& {Ballester}, J.~L. 2009{\natexlab{a}}, \apj, 707,
  662

\bibitem[{{Soler} {et~al.}(2009{\natexlab{b}}){Soler}, {Oliver}, {Ballester},
  \& {Goossens}}]{Soler.etal2009a}
{Soler}, R., {Oliver}, R., {Ballester}, J.~L., \& {Goossens}, M.
  2009{\natexlab{b}}, \apjl, 695, L166

\bibitem[{{Sterling} \& {Moore}(2005)}]{Sterling.Moore2005}
{Sterling}, A.~C., \& {Moore}, R.~L. 2005, \apj, 630, 1148

\bibitem[{{Sterling} {et~al.}(2001){Sterling}, {Moore}, \&
  {Thompson}}]{Sterling.etal2001}
{Sterling}, A.~C., {Moore}, R.~L., \& {Thompson}, B.~J. 2001, \apjl, 561, L219

\bibitem[{{Su} {et~al.}(2012){Su}, {Wang}, {Veronig}, {Temmer}, \&
  {Gan}}]{Su.etal2012a}
{Su}, Y., {Wang}, T., {Veronig}, A., {Temmer}, M., \& {Gan}, W. 2012, \apjl,
  756, L41

\bibitem[{{Tandberg-Hanssen}(1995)}]{Tandberg-Hanssen1995}
{Tandberg-Hanssen}, E., ed. 1995, Astrophysics and Space Science Library, Vol.
  199, {The Nature of Solar Prominences}

\bibitem[{{Vial} {et~al.}(2012){Vial}, {Olivier}, {Philippon}, {Vourlidas}, \&
  {Yurchyshyn}}]{Vial.etal2012}
{Vial}, J.-C., {Olivier}, K., {Philippon}, A.~A., {Vourlidas}, A., \&
  {Yurchyshyn}, V. 2012, \aap, 541, A108

\bibitem[{{Wang} \& {Muglach}(2013)}]{Wang.Muglach2013}
{Wang}, Y.-M., \& {Muglach}, K. 2013, \apj, 763, 97

\bibitem[{{Wang} {et~al.}(2009){Wang}, {Muglach}, \& {Kliem}}]{Wang.etal2009}
{Wang}, Y.-M., {Muglach}, K., \& {Kliem}, B. 2009, \apj, 699, 133

\bibitem[{{Wedemeyer} {et~al.}(2013){Wedemeyer}, {Scullion}, {Rouppe van der
  Voort}, {Bosnjak}, \& {Antolin}}]{Wedemeyer.etal2013}
{Wedemeyer}, S., {Scullion}, E., {Rouppe van der Voort}, L., {Bosnjak}, A., \&
  {Antolin}, P. 2013, \apj, 774, 123

\bibitem[{{Zirker} {et~al.}(1998){Zirker}, {Engvold}, \&
  {Martin}}]{Zirker.etal1998}
{Zirker}, J.~B., {Engvold}, O., \& {Martin}, S.~F. 1998, \nat, 396, 440

\end{thebibliography}


\end{document}